\documentclass[prl,nofootinbib,twocolumn,showpacs,preprintnumbers,amsmath,amssymb,floatfix,superscriptaddress]{revtex4}
\usepackage{amsmath}
\usepackage{amsfonts}
\usepackage{dcolumn}                    	     	
\usepackage{amssymb}
\usepackage{graphicx,graphics,rotating}     		
\usepackage[english]{babel}	
\usepackage{bm,fancybox}                	     	

\begin{document}

\title{Localization of resonance eigenfunctions on quantum repellers }
\author{Leonardo Ermann}
\affiliation{Departamento de F\'\i sica, CNEA, Libertador 8250, (C1429BNP) Buenos Aires, Argentina}
\affiliation{Departamento de F\'\i sica, FCEyN, UBA, Pabell\'on 1 Ciudad 
Universitaria, C1428EGA Buenos Aires, Argentina}
\author{Gabriel G. Carlo} 
\affiliation{Departamento de F\'\i sica, CNEA, Libertador 8250, (C1429BNP) Buenos Aires, Argentina}
\author{Marcos Saraceno} 
\affiliation{Departamento de F\'\i sica, CNEA, Libertador 8250, (C1429BNP) Buenos Aires, Argentina}
\affiliation{Escuela de Ciencia y Tecnolog\'\i a, UNSAM, Alem 3901, B1653HIM Villa Ballester, Argentina}

\date{\today}

\pacs{05.45.Mt, 03.65.Sq, 03.65.Yz}

\begin{abstract}
We introduce a new phase space representation for open 
quantum systems. This is a very powerful tool to help advance in the study of the 
morphology of their eigenstates. We apply it to two different versions of a paradigmatic model, 
the baker map. This allows to show that the long-lived resonances 
are strongly scarred along the shortest periodic orbits that belong to the classical repeller. 
Moreover, the shape of the short-lived eigenstates is also analyzed. Finally, 
we apply an antiunitary symmetry measure to the resonances that permits to quantify 
their localization on the repeller.
\end{abstract}

\maketitle

Open quantum systems with classical chaotic behavior
have been far less explored than their closed counterparts, 
and as a consequence several of their properties are less known. 
This situation is in contradiction with their importance 
in many areas of physics, some of them of very recent development. 
One of the most significant examples is the study of quantum to classical correspondence 
\cite{Quantum2Classical}. But their applicability 
range extends to microlasers \cite{Microlasers,Microlasers2,WiersigPrl06}, 
quantum dots \cite{QuantumDots}, chaotic scattering 
\cite{ChaoticScattering}, and more topics. The lack of unitarity at the quantum level, 
and the complex - and mostly fractal - nature of the repeller sets is at the heart 
of the difficulties in their understanding. 

The classical phase space of open chaotic systems is characterized 
by the forward and backward trapped sets, i.e., fractal sets of trajectories that stay 
trapped forever in the future and in the past, respectively. The intersection 
of both sets is called the repeller. Classical properties of this set are its decay rate 
$\gamma_{cl}$, various generalized dimensions and the spectral properties of the 
Perron-Frobenius operator \cite{ChaoticScattering}. On the other hand, the quantum evolution is described 
by a nonunitary operator, whose right and left eigenstates (the so called resonances) 
are nonorthogonal and the eigenvalues $\lambda_{i}$ are complex with modulus less than or equal to one
($\nu_i^{2}=\vert\lambda_{i}\vert^{2}=\exp(-\Gamma_{i})\leq1$, where $\Gamma_{i}\geq0$ is the decay rate).
The short-lived resonances ($\Gamma_{i} \gg1$) are associated with the trajectories that escape 
from the system before the Ehrenfest time, while the long-lived ones ($\Gamma_{i}= \mathcal O(1)$) 
are related to the classical forward or backward trapped sets 
(in \cite{ShepelyanskyPhD99} they were coined quantum 
fractal eigenstates). According to the conjectured fractal Weyl 
law, the number of long-lived resonances $N_{\gamma}$ can be related to $\hbar$ and the 
structure of the classical phase space through the expression $N_{\gamma}\sim\hbar^{-(d-1)}$ 
($d$ is the fractal dimension of the classical strange repeller). This law has been checked for 
a three disk system \cite{Lu} and some quantum maps 
\cite{Shepelyansky,Schomerus,Nonnenmacher1,Nonnenmacher2,Keating,Wisniacki,Pedrosa}. 

Finer information about the semiclassical limit is provided by the structure of eigenfunctions, 
about which much less is known. For the open baker map it was found that the probability 
density averaged for several right eigenstates 
is supported by the corresponding classical forward trapped set \cite{Keating}. A simple 
argument by analogy to Shnirelman's theorem would favor a weak equidistribution of the 
single spectral components over the repeller, but of course this leaves ample 
room for ``accidental'' localization on the periodic orbits, 
{\it i.e.} for the existence of scars in open quantum systems. 
Such long-lived scarred modes have been observed in optical microcavities 
\cite{WiersigPrl06}, and have been extensively studied in closed systems \cite{HellerPrl84}. 
Recently, some results for the open cat map 
have been provided in \cite{Wisniacki}. But the knowledge we have 
about this behavior is still rudimentary.

In this letter we show that many long lived resonances of the open baker map, 
besides the obvious localization on the repeller, indeed show an enhancement on some 
of its shortest periodic orbits. To this purpose we introduce a phase space representation 
of the spectral components of the open map that is in complete analogy to the usual Husimi 
functions of the closed case. We also investigate the shape of short-lived resonances 
of the open triadic baker map. We find that structures resembling the states of a 
simple Hamiltonian develop outside of the classical repeller.
On the other hand, we have applied a measure of antiunitary 
symmetry to the eigenstates. By doing this we were able to quantify the localization 
of a given resonance on the repeller. Moreover, this measure turns out to be simply related to 
the moduli of the eigenvalues.
We believe that these tools could provide an alternative way of identifying long and short-lived
resonances, a task of the utmost importance in the road to demonstrate 
the conjectured Weyl law \cite{Schomerus}. 
In fact, this stands as an open problem, where only for the Walsh quantized 
open baker map \cite{Nonnenmacher2,nonn-keating} exact results could be obtained. 

We consider a closed quantum map $U$, chosen as the quantization of a classical map which 
is chaotic. The phase space is the unit torus and upon quantization we impose antiperiodic 
boundary conditions resulting in a Hilbert space of dimension $N$, 
where $\hbar=(2\pi N)^{-1}$ plays the role of an effective Planck's constant.
$U$ is then an $N\times N$ unitary matrix.
Our maps have parity ($R$) and time reversal ($T$) symmetries in both its classical and 
quantum versions. 

The map is ``opened'' as usual by means of a projection operator $\Pi_o$ that defines the 
escape region in phase space. In this way, $\Pi=1-\Pi_o$ composed with the closed map defines 
the open system. Depending on the symmetries of 
the projector it is possible to maintain or break in the open map the symmetries of the closed one. 
The resulting operator, $\tilde U=U\Pi$ is not normal and has a null subspace whose 
dimension is $ tr(\Pi_o)$. 
If we assume that it is diagonalizable, its spectral decomposition
 in terms of right and left eigenstates is
\begin{equation}\label{eq:Udecomp}
\tilde{U}=\sum_i\lambda_i \frac{\vert\psi^R_i\rangle\langle\psi^L_i\vert}
{\langle\psi^L_i\vert\psi^R_i\rangle}
\end{equation}
where we assume the normalization condition $\langle\psi^{L}_i\vert\psi^{L}_i\rangle=
\langle\psi^{R}_i\vert\psi^{R}_i\rangle=1$. 
As the map is contracting, the spectrum $\lambda_i$ is strictly inside the unit circle.
Using the decomposition in Eq.\ref{eq:Udecomp} the autocorrelation of a coherent state 
$ \vert q,p\rangle$ can be written as
\begin{equation}\label{eq:autocorrel}
\langle q,p\vert \tilde U^n\vert q,p\rangle=\sum_i\lambda_i^n h_i(q,p)
\end{equation}
in terms of complex phase space distributions 
\begin{equation}
h_i(q,p)\equiv\frac{\langle q,p\vert\psi_i^R\rangle\langle\psi_i^L\vert q,p\rangle}
{\langle\psi^L_i\vert\psi^R_i\rangle}
\end{equation}
As the right (left) eigenfunctions are approximately concentrated
on the backwards (forward) trapped sets, we expect $\vert h_i(p,q)\vert$ to map out the 
distribution of the single resonances on the intersection of these sets, {\it i.e.} on the 
repeller. Thus they constitute the ideal tool to study the localization properties and 
eventually the  ``scarring'' in open systems. 
Notice that $\vert h_i(p,q)\vert\propto\sqrt{{\mathcal H}^R_i(p,q)~{\mathcal H}^L_i(p,q)}$,
where ${\mathcal H}^{R,L}_i$ are Husimi functions of the $R,L$ eigenstates, and that for 
unitary systems they reduce to the Husimi functions of the eigenstates.
We would like to mention that, instead of considering coherent states we could have chosen any other 
representation in phase space. Different choices can help to analyze different properties of ${\vert\psi_i^R\rangle\langle\psi_i^L\vert}/{\langle\psi^L_i\vert\psi^R_i\rangle}$.
\begin{figure}[t!]
\begin{center}
 \includegraphics[width=0.48\textwidth]{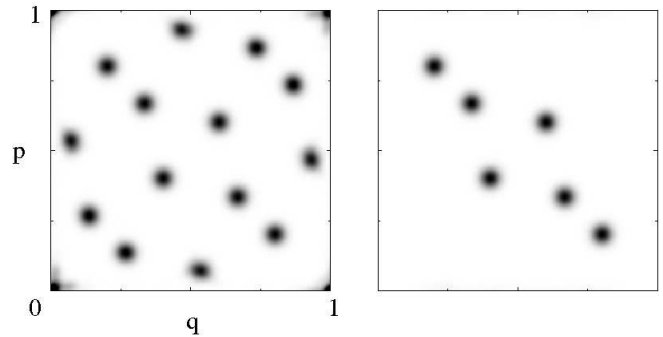}
\caption{Autocorrelation function in phase space for the $4$th iteration of the standard quantum 
baker map for $N=160$. The closed version, showing all periodic points of period 
$4$ is on the left panel, and the open version with $l=3$ is on the right panel. 
The periodic orbits which have the forbidden sequence ($000$ and $111$) are absent 
in the open system.}\label{fig:correl}
\end{center}
\end{figure}

We apply these ideas to study the localization of resonance eigenfunctions for the open baker map. 
In its standard version it is defined as $(p^\prime=p/2+\lfloor2q\rfloor/2,q^\prime=2q-\lfloor2q\rfloor$), 
where $\lfloor2q\rfloor$ is the integer part. It has a complete symbolic dynamics in terms of two symbols. 
The quantization  is obtained in terms of the anti--periodic, discrete 
Fourier transform $\left(G_N\right)_{j,k}\equiv exp{(-i2\pi/N(j+1/2)(k+1/2)}/
\sqrt{N}$, as $U=G_N^\dagger \ diag{(G_{N/2},G_{N/2})}$ \cite{voros,saraceno}. 
The advantage for this kind of study is that one can easily engineer openings
that coincide with some iteration of the basic Markov partition and thus obtain repellers 
with a simple symbolic description in terms of a subshift of finite type. For concreteness we take 
the escape region $\Pi_o$ to be the union of the areas defined by 
$q_1\in[0,1/2^l]\times p_1\in[0,1]$ and $q_2\in[1-1/2^l,1]\times p_2\in[0,1]$ leading to a subshift 
for which the complete symbolic dynamics is pruned of all trajectories that include the 
strings $0^l$ and $1^l$. 
The opening is symmetric with respect to $q=1/2$ in order to retain the parity symmetry.
We will focus on the first non-trivial case of $l=3$ where $q_1\in[0,1/8]$ $q_2\in[7/8,1]$, 
and the strings $000$ and $111$ are pruned. 
The topological entropy $\lambda_T\approx 0.4812$ is easily calculated from the leading 
eigenvalue of the transition matrix. 
In Fig. \ref{fig:correl} we compare the autocorrelation function in phase space (Eq. \ref{eq:autocorrel}) 
for both the open and the closed versions of the fourth iteration of the map clearly showing 
the disappearance of the periodic points forbidden by the pruning rules.

\begin{figure}[t!]
\begin{center}
 \includegraphics[width=0.48\textwidth]{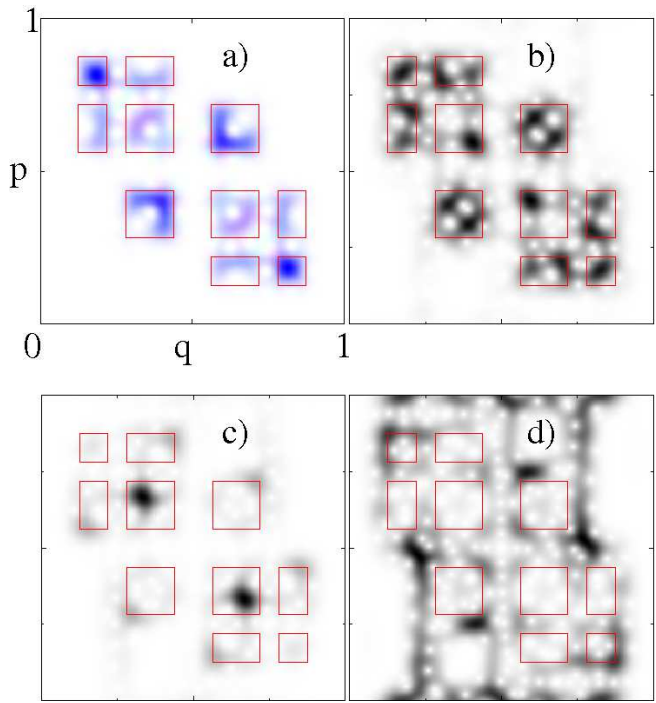}
\caption{(color online) Phase space representation of a set of resonances 
for the standard open baker map with $N=320$ and $l=3$. 
In panel $a)$ we display $h_i(q,p)$ for the longest-lived resonance ($\vert\lambda_{a}\vert\approx0.977$). 
Complex phase is represented by color ranging from blue ($0$) to red ($2 \pi$). 
In $b)$, $c)$ and $d)$ we show $\left\vert h_i(q,p)\right\vert$ for eigenstates corresponding to 
$\vert\lambda_{b}\vert\approx0.905$, $\vert\lambda_{c}\vert\approx0.856$, and $\vert\lambda_{d)}\vert\approx0.641$, 
respectively. The classical repeller for times between $t=-2$ and $t=5$ lies inside the solid red 
line rectangles.}
\label{fig:HusOpBakL3}
\end{center}
\end{figure}

Four examples of resonance eigenfunctions are displayed in Fig.\ref{fig:HusOpBakL3}, using the distribution  
$h_i(q,p)$. They correspond to a Hilbert space of dimension $N=320$ and $l=3$. 
In panel $a)$ we show the longest-lived resonance for which $\vert\lambda_{a}\vert\approx0.977$. 
In this case, the complex function $h(q,p)$ is represented 
through an intensity (moduli) and color (phase) scale, this latter ranging from blue ($0$) to red ($2 \pi$). 
It is clear that its values are almost completely real, this being an expected property of an 
eigenstate whose associated eigenvalue is so close to the unit circle. 
In panels $b)$, $c)$ and $d)$ we show $\left\vert h_i(q,p)\right\vert$ for eigenstates corresponding to 
$\vert\lambda_{b}\vert\approx0.905$, $\vert\lambda_{c}\vert\approx0.856$, and 
$\vert\lambda_{d)}\vert\approx0.641$ respectively (in these cases we 
restrict our analysis to the moduli of the distribution). 
We now focus on the specific shape of resonances. Those in $a)$ and $c)$ clearly 
show an enhanced probability on the periodic orbits $0011$ and $01$ respectively, which belong to the repeller. 
This enhancement on the periodic points and their manifolds is typical of the scarring phenomenon. In contrast, 
$b)$ has no visible prevailing orbit in its structure, and it is distributed almost uniformly on the repeller. 
We have represented a finite time version of the classical repeller by means of red line rectangles as a guide 
to the eye. We have taken times between $t=-2$ and $t=5$ to keep symmetries intact. Finally, in order to 
complete the picture of a representative set of resonances, we include one that it is localized out 
of the repeller, in panel $d)$.

In order to study the eigenstates outside of the repeller in a clear way, we choose to analyze 
the triadic baker map $\mathcal{B}_T$. This map is defined as 
$(p^\prime=p/3,q^\prime=3q)$ if $0 \leq q < 1/3$, $(p^\prime=(p+1)/3,q^\prime=3q-1)$ if $1/3 \leq q < 2/3$, 
and $(p^\prime=(p+2)/3,q^\prime=3q-2)$ if $2/3 \leq q < 1$. From our point of view, 
the main advantage of this map is that it has an isolated fixed point at $(q,p)=(1/2,1/2)$, 
far from the discontinuities.
\begin{figure}[t!]
\begin{center}
 \includegraphics[width=0.48\textwidth]{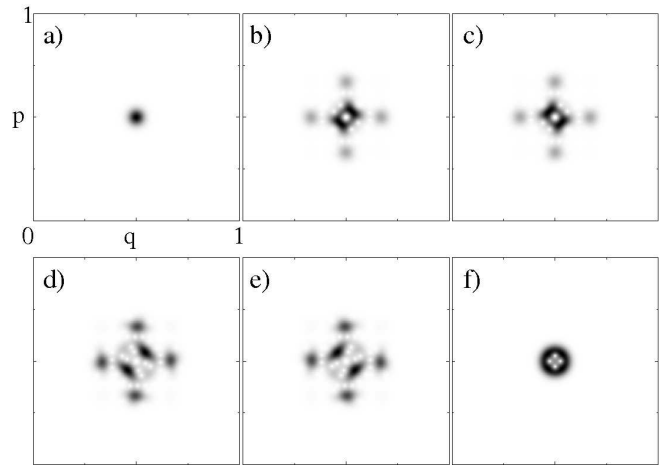}
\caption{(color online) Phase space representation $\left\vert h_i(q,p)\right\vert$ 
of the six longest-lived resonances 
for the triadic open baker map with $N=243$. Moduli of the corresponding eigenvalues 
are $\vert\lambda_{a}\vert=1/\sqrt{3}\approx0.577$, $\vert\lambda_{b}\vert=\vert\lambda_{c}\vert\approx0.249$,  $\vert\lambda_{d}\vert=\vert\lambda_{e}\vert\approx0.184$ and $\vert\lambda_{f}\vert\approx0.064$. 
The map is real, therefore $\lambda_{a},\lambda_{f}\in\mathbb{R}$,   $\lambda_{b}=\lambda_{c}^{\ast}$, and $\lambda_{d}=\lambda_{e}^{\ast}$.}
\label{fig:HusOpBakCent}
\end{center}
\end{figure}
The quantization of this map in a $N$--dimensional Hilbert space ($N=2\pi/\hbar$, divisible by $3$) 
with antisymmetric boundary conditions is given by 
\begin{equation}
B_T=G^\dagger_N\left(\begin{array}{ccc}G_{N/3}&0&0\\
0&G_{N/3}&0\\0&0&G_{N/3}\end{array}\right) \nonumber
\end{equation}
The open version of the map can be obtained in the same way 
as before, by means of a projector $\Pi$, but this time on the central region in position.
This amounts to saying that in the quantum case the left and right thirds of the 
columns of $B_T$ are set to zero. This opens two thirds of the phase space 
(where the forbidden region is  given by $q\in[0,1/3]$ and $q\in[2/3,1]$, 
satisfying  $tr(\Pi)=N/3$ in the quantum case). 
The repeller of this system is reduced to the point $(q,p)=(1/2,1/2)$. It is important to underline 
that this map is hyperbolic, but it is not chaotic. However, it is very suitable in order 
to unveil the structure 
of resonances lying outside of the repeller, that in the previously studied version of the baker map 
have a complicated structure. 
Fig. \ref{fig:HusOpBakCent} shows the six longest-lived resonances for $N=3^{5}$. 
In this case the morphology is very clear. They resemble excited 
states of the hyperbolic Hamiltonian $H=qp$, centered at the fixed point $(q,p)=(1/2,1/2)$. 
This suggests that the longest-lived fraction of the short-lived resonances associated to 
more complicated repellers could be 
interpreted (and perhaps theoretically constructed) as convenient superpositions of these simpler 
structures. This completes a very interesting picture, i.e., one could try to devise a theory of 
short periodic orbits capable of describing the complete significant quantum information of an open system 
\cite{future}. 

\begin{figure}[t!]
\begin{center}
 \includegraphics[width=0.48\textwidth]{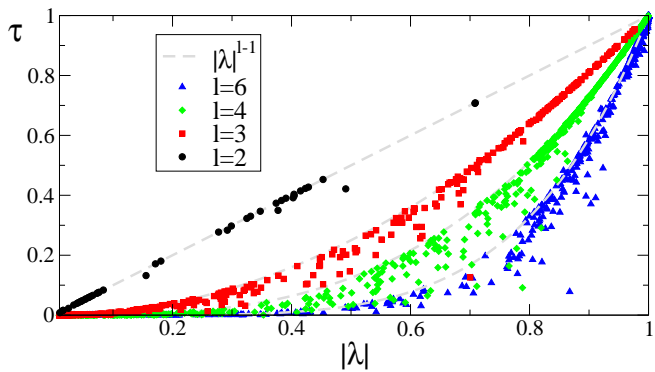}
\caption{(color online)
Measure of time reversibility of a given eigenstate 
($\tau_i\equiv\vert\langle\psi^R_i\vert\hat{T}\vert\psi^L_i\rangle\vert$) as a function of the 
modulus of its eigenvalue, for the standard open baker map with $l=2,3,4,6$ and $N=1024$. 
Dashed (gray) lines correspond to $\vert\lambda^{l-1}\vert$.}
\label{fig:TRvsMod}
\end{center}
\end{figure}

Finally, the time reversal symmetry provides us with very interesting results 
regarding the classification of resonances into long and short-lived. In the classical 
case this symmetry can be restricted to the orbits belonging to the repeller, 
since they are not affected by the presence of the escape regions, in our example. 
In other words, the repeller corresponding to our open system retains the time 
reversal invariance of the closed one. However, 
if we look at resonances in Fig.\ref{fig:HusOpBakL3} we immediately notice that 
in general they are not symmetric with respect to the line $q=p$, this 
indicating that they are not time reversal invariant (in general left eigenstates 
cannot be obtained from applying the time reversal operator to the right ones, 
for example). Then, for the quantum case we can define a measure of the time reversal 
symmetry for each eigenstate in order to see if this is related to its localization 
on the repeller. This is accomplished by $\tau_i\equiv\langle\psi^R_i\vert T\vert\psi^L_i\rangle$, 
where $T=KG$ ($K$ stands for the complex conjugation operator). In the case of 
maps having a time reversal symmetry $\tau_i=1$ for any $i$. 
Results for the standard open baker map can be seen in Fig. \ref{fig:TRvsMod}. 
We have found that there is a simple relation among the values of this measure and the 
eigenvalues moduli, given by $\tau_i=\vert\lambda\vert^{l-1}$. 
On one hand this confirms that the resonances that have a maximum overlap with the 
repeller (long-lived ones) show the higher time reversibility. On the other hand, 
it confirms this quantity as an alternative way to distinguish between long and 
short-lived resonances. We think that by following this approach a prediction on 
the number of each kind of eigenstates could be found \cite{future}.

In conclusion, we have introduced a very powerful phase space representation for 
open quantum systems that is in complete analogy with the usual Husimi distribution 
for closed ones. This allowed us to clearly show that long-lived resonances can 
be strongly scarred along the shortest periodic orbits that belong to the corresponding 
classical repeller. Moreover, we have obtained some hints on the structure of the short-lived 
eigenstates, i.e. those living outside of the invariant set. We believe that short periodic 
orbits of the repeller play an important role also in this case. Finally, a time 
reversibility measure for the resonances has been devised. We were able to find a 
simple relation with the eigenvalues moduli showing that the long-lived resonances 
have the highest time reversibility. We consider that these tools are very promising 
in the way to demonstrate the fractal Weyl law for open quantum systems.

\begin{acknowledgments}
Partial support by CONICET and ANPCyT is gratefully acknowledged.
\end{acknowledgments}


\begin{thebibliography}{99}

\bibitem{Quantum2Classical}
W.H. Zurek and J.P. Paz, Phys. Rev. Lett. \textbf{72}, 2508 (1994).

\bibitem{Microlasers}
W. Fang, Phys. Rev. A {\bf 72}, 023815 (2005); J.U. N\"ockel and 
D.A. Stone, Nature (London) {\bf 385}, 45 (1997); 
T. Harayama, P. Davis and K.S. Ikeda,
Phys. Rev. Lett. {\bf 90}, 063901 (2003);

\bibitem{Microlasers2}
J. Wiersig and M. Hentschel, Phys. Rev. A {\bf 73} 031802(R) (2006); 
Phys. Rev. Lett. \textbf{100}, 033901 (2008).
 
\bibitem{WiersigPrl06}
J. Wiersig, Phys. Rev. Lett. {\bf 97} 253901 (2006).

\bibitem{QuantumDots}
R. Akis {\em et al.}, Phys. Rev. Lett. {\bf 79}, 123 (1997).

\bibitem{ChaoticScattering}
P. Gaspard, \textit{Chaos, Scattering and Statistical Mechanics}, 
Cambridge Univ. Press, Cambridge (1998); 
C. Jung and T.H. Seligman, Phys. Rep. {\bf 285}, 77 (1997). 

\bibitem{ShepelyanskyPhD99} 
G. Casati, G. Maspero, and D.L. Shepelyansky, 
Physica D {\bf 131} 311 (1999).

\bibitem{Lu} 
W.T. Lu, S. Sridar and M. Zworski, 
Phys. Rev. Lett \textbf{91}, 154101 (2003).

\bibitem{Shepelyansky} 
D.L. Shepelyansky, Phys. Rev. E \textbf{77}, 15202(R) (2008).

\bibitem{Schomerus} 
H. Schomerus and J. Tworzydlo, Phys. Rev. Lett. \textbf{93}, 154102 (2004).

\bibitem{Nonnenmacher1} S. Nonnenmacher and M. Rubin, Nonlinearity \textbf{20}, 
1387 (2007).

\bibitem{Nonnenmacher2} S. Nonnenmacher and M. Zworski, 
J. Phys. A: Math. Gen. \textbf{38}, 10683 (2005).

\bibitem{Keating} 
J.P. Keating, M. Novaes, S.D. Prado and M. Sieber, 
Phys. Rev. Lett. \textbf{97}, 150406 (2006).

\bibitem{Wisniacki} D. Wisniacki and G.G. Carlo, 
Phys. Rev. E \textbf{77}, 45201(R) (2008).

\bibitem{Pedrosa} J.M. Pedrosa, G.G. Carlo, D.A. Wisniacki and L. Ermann, 
Phys. Rev. E \textbf{79}, 016215 (2009).

\bibitem{HellerPrl84}
E.J. Heller, Phys. Rev. Lett. {\bf 53} 1515 (1984).

\bibitem{nonn-keating}
J.P. Keating, S. Nonnenmacher, M. Novaes and M. Sieber, 
Nonlinearity \textbf{21}, 2591 (2008).

\bibitem{voros} 
N.L. Balazs and A. Voros, Ann. Phys. \textbf{190}, 1 (1989).

\bibitem{saraceno} 
M. Saraceno, Ann. Phys. \textbf{199}, 37 (1990).

\bibitem{future}
L. Ermann, G.G. Carlo and M. Saraceno, in preparation.

\end{thebibliography}
\end{document}